\def\JPCM{{J. Phys. Condens. Matter\ }\/}
\def\JMMM{{J. Magn. Magn. Mater.\ }\/}
\def\PR{{Phys. Rev.\ }\/ }

\def\ber{\begin{eqnarray}}
\def\eer{\end{eqnarray}}
\def\bers{\begin{eqnarray*}}
\def\eers{\end{eqnarray*}}
\def\be{\begin{equation}}
\def\ee{\end{equation}}

\documentclass[prb,aps,showpacs,superscriptaddress,epsfig,twocolumn,floatfix]{revtex4}
\usepackage{dcolumn}
\usepackage{longtable}

\usepackage{graphicx}
\usepackage{color}
\begin{document}

\title{Signature effects of spin clustering and distribution of spin couplings \\
on magnetization behaviour in Ni-Fe-Mo and Ni-Fe-W alloys}
\author{Mitali Banerjee}
\affiliation{Department of Materials Science, S.N. Bose National Centre for Basic Sciences,
JD Block, Sector III, Salt Lake City, Kolkata 700098, India}
\author{Avinash Singh}
\affiliation{Department of Physics, Indian Institute of Technology, Kanpur 208016, India}
\author{A. K. Majumdar}
\email{akm@bose.res.in}
\affiliation{Department of Materials Science, S.N. Bose National Centre for Basic Sciences, 
JD Block, Sector III, Salt Lake City, Kolkata 700098, India}
\author{A. K. Nigam}
\affiliation{Department of Condensed Matter Physics and Materials Science, Tata Institute of Fundamental Research, Homi Bhabha Road, Colaba, Mumbai 400005, India}
\date{\today}

\begin{abstract}
The spontaneous magnetization as a function of temperature is investigated for a number of disordered Ni-Fe-Mo and Ni-Fe-W alloys using superconducting quantum interference device magnetometry, with a focus on the low-$T$ behavior as well as the critical exponents associated with the magnetic phase transition. While the low-$T$ magnetization is found to be well described by Bloch's $T^{3/2}$ law, an extraordinary enhancement of the spin-wave parameter $B$ and the reduced coefficient $B_{3/2}$=$BT_{\rm C}^{3/2}$ are observed with increasing Fe dilution as compared to conventional 3d ferromagnets, whereas the critical amplitudes are found to decrease systematically. Recent locally self-consistent calculations of finite-temperature spin dynamics in a generic diluted magnet provide an understanding in terms of two distinct energy scales associated with weakly coupled bulk spins in the FM matrix and strongly coupled cluster spins. In view of similar behaviour observed in diluted magnetic semiconductors and other ferromagnetic alloys, it is proposed that these distinctive features corresponding to the three important temperature regimes provide macroscopic indicators of signature effects of spin clustering on magnetization behaviour in disordered ferromagnets. 

\end{abstract} 
\maketitle

\section{Introduction}
Magnetic phases of alloy systems with both ferromagnetic and antiferromagnetic exchange interactions have been studied in details by many groups including ours. Ni-rich Ni-Fe-Cr and Ni-Fe-V alloys\cite{1,2} have shown exotic magnetic phases and hence have been probed extensively. Compositions near Ni80\% - Fe20\% have been widely investigated in bulk form not only for their rich magnetic phases but also for the unique quality of low magnetostriction and high initial permeability\cite{3}. These are ideal properties for making magnetic cores for light electrical equipment as well as magnetic shielding materials. 

The observation of several signatures of a spin-glass-like phase in recent dc magnetization and ac susceptibility studies of Ni-Fe-Mo and Ni-Fe-W alloys\cite{4,5,6} indicates presence of competing ferromagnetic and antiferromagnetic spin interactions, resulting in frustration-induced frozen spin configurations in spin clusters. Generally, a broad distribution of magnetic spin couplings would be expected corresponding to different separations and configurations resulting from the positional disorder of magnetically active Fe atoms in these alloys. 

Are there any distinctive finite-temperature spin dynamics effects associated with this disorder-induced distribution of magnetic interactions in the Ni-Fe-Mo and Ni-Fe-W alloys which are observable in their macroscopic magnetization behaviour? A similar question was recently addressed in the context of squid magnetization studies\cite{7} of diluted magnetic semiconductors Ga$_{1-x}$Mn$_{x}$As, which also exhibit strong disorder effects due to Mn positional disorder. Recent  theoretical studies of finite-temperature local spin dynamics within a generic model for diluted magnets have provided fundamental understanding of macroscopic magnetization characteristics in terms of microscopic spin disorder, spin clustering, and distribution of spin couplings.\cite{8,9,10}

There are three important temperature regimes in the magnetization behaviour of a generic disordered ferromagnet. The low-temperature regime characterized by the spin-wave parameter, the intermediate-temperature regime nearly upto $T_{\rm C}$
characterized by the overall shape of the magnetization decay, and the critical regime very close to $T_{\rm C}$ involving critical fluctuations, divergent spin correlation length, and characterized by the critical exponents and critical
amplitudes. In this paper we propose that the magnetization behaviours in all three temperature regimes actually provide macroscopic indicators of signature effects associated with microscopic spin dynamics of weakly coupled bulk spins and strongly coupled cluster spins. For that purpose we have made a detailed study of the temperature dependence of magnetization in Ni-Fe-Mo and Ni-Fe-W alloys and the critical exponents and critical amplitudes associated with the phase transition.

\section{Experimental details}
Alloys of both ternary series (Mo and W) were prepared by arc melting of the required amount of constituents of Johnson-Mathey, 99.999 $\%$ pure Ni, Fe, Mo, and W in pure argon atmosphere. Then the shiny buttons were sealed in quartz tubes, flushed with ultra pure argon gas, and homogenized at  1300$^{o}$C for 48 hours. Then they were cold-rolled and cut to size. A final annealing was done in argon atmosphere at 1100$^{o}$C to reduce strain due to cold rolling. We kept the concentration of Ni roughly around 80\ - 83 at. $\%$ and varied Fe from 17 to 3 at. $\%$ and Mo from 2 to 14 at. $\%$. For alloys containing W, Ni concentration remains roughly around 79 to 86 at. $\%$ while that of Fe steadily decreases from 18 to 3 at. $\%$ and that of W increases from 3 to 10 at. $\%$. All magnetic measurements were done using Quantum Design SQUID magnetometer (QDMPMS). The Curie temperatures ($T_{\rm C}$) were determined from the rate of fall of magnetization as a function of temperature in applied fields of ~ 20\ - 50 Oe. The compositions of the samples have been confirmed by energy dispersive spectrum (EDS) measurements using a scanning electron microscope (SEM). The details of measurements of other quantities will be discussed along with the presentation of data in section III.

\section{Results and discussion}

\subsection*{A.	Magnetic phase of the alloys}
The X-ray diffraction pattern reveals that all the alloys are of single FCC phase with no additional phase. The lattice parameter varies by 1.5-2 $\%$ from that of pure Ni and increases with the increase of Mo/W. The alloys of both the Mo and W series show paramagnetic to ferromagnetic transition at $T_{\rm C}$, which varies widely depending on the amount of Fe present in the alloy. The magnetization of the alloys with low Fe content are measured in $\sim$ 20 Oe applied field and with low temperature attachment of QDMPMS and that for the alloys with higher concentration of Fe were measured in $\sim$ 50 Oe applied field with high temperature attachment of QDMPMS. The $T_{\rm C}$ for all the compositions are already reported by Banerjee et al.\cite{4,5} and are given in Table I. It is clear that increasing Mo and W content and accompanying decreasing Fe content in the alloys lowers the $T_{\rm C}$ rapidly. 

In the Mo series, the sample Mo13.5 shows a second phase transition apart from the paramagnetic-ferromagnetic transition found in all the alloys of the series. No such additional phase has been found in the W series which we anticipate is due to lack of high W samples around the critical concentration. We have found that the $T_{\rm C}$ extrapolates to 0 K for $\sim$14.2 at. $\%$ Mo whereas the Slater-Pauling curve\cite{11} gives $T_{\rm C}$ $\sim$ 0 K for 18 $\%$ Cr in Ni. The split-band model of Berger\cite{12}, however, gives the same critical concentration of $\sim$ 12 $\%$ for Cr, Mo, and W since they fall in the same group in the periodic table. Detailed magnetic measurements, both dc and ac, have been done on Mo13.5 to establish that the second phase is indeed a mixed ferro-spin-glass phase. The $M(H)$ measurements show a ferromagnetic nature with a feeble non-saturating character at 5 K but grossly it is ferromagnetic. The peak in ac susceptibility shifts to higher temperatures with increasing frequency and also the field cooled (FC) and zero-field cooled (ZFC) curves bifurcate until an applied field $\sim$ 100 Oe. These characteristic features resemble those of spin glasses. All the above will be presented in detail elsewhere\cite{6}. 

The alloys of both the series show very thin $M$-$H$ loops just like their parent permalloy. However, the permeability found is not comparable to those of permalloys. Magnetic annealing of the alloys could give higher permeability\cite{3}. The saturation magnetization falls from 0.85 $\mu_B$ to 0.08 $\mu_B$ on increment of Mo in the alloys\cite{4} while it decreases from 0.85 $\mu_B$ to 0.11 $\mu_B$ for the W series\cite{5}. The ac-susceptibility measurements show the same $T_{\rm C}$ as found from dc measurements.

\subsection*{B.	Low-temperature high-field magnetization of the alloys}
In conventional ferromagnets, the temperature dependence of the spontaneous magnetization $M_s (T)$ far below $T_{\rm C}$ is dominated by long-wavelength spin-wave excitations. The excitation energy $E(k)$ of spin waves in the limit of small wave vectors ($k$$\ll$$a^{-1}$, $a$ is the lattice spacing) is given by
\begin{equation} \label{disp}
E(k)= \hbar\omega(k)= g\mu_B H_{\mathrm{int}} + Dk^2 + Ek^4 + \ldots , 
\end{equation}
where the first term is an energy gap due to the presence of an effective internal field $H_\mathrm{int}$, arising from the applied field, the anisotropy field, and the spin-wave demagnetizing field. $D$ is the spin-wave stiffness constant, and $E$ is a proportionality constant. Even in disordered ferromagnets, there is ample experimental evidence that long-wavelength spin-wave modes are a useful way to represent the low-energy magnetic excitations.\cite{kaul_1983} 

\begin{figure} [b]
\centering
\includegraphics[width=10cm,height=8.5cm]{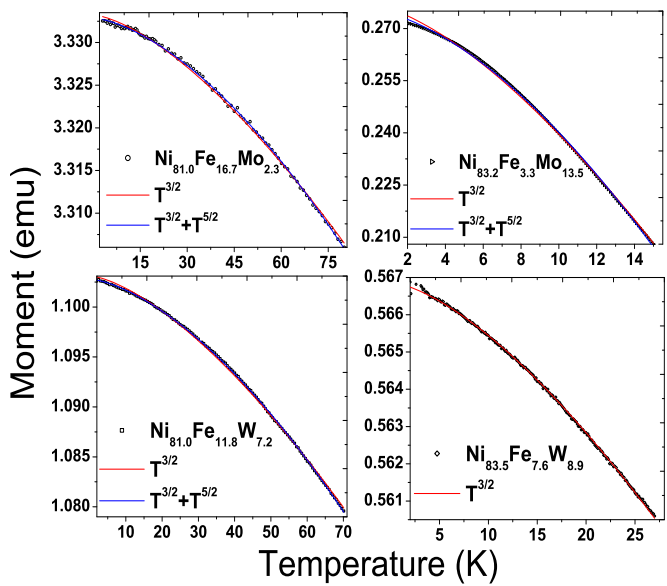}
\caption{(Color Online) Magnetisation vs. Temperature plot for four samples are given, among them three samples fit better using Eq. (\ref{spin}) including anharmonic term another one shown fitted to Bloch's T$^{3/2}$ law.}
\label{fig1}
\end{figure}

In the low-temperature limit, according to the Heisenberg model, the change in the spontaneous magnetization due to the excitation of spin waves can be written as:\cite{13}
\begin{eqnarray} \label{spin}
M(T) & = & M(0)[1 - Bz\left(\frac{3}{2},\frac{T_g}{T}\right)T^{3/2}\nonumber\\
& - & Cz\left(\frac{5}{2},\frac{T_g}{T}\right)T^{5/2}+\ldots], 
\end{eqnarray}
where $M(0)$ is the magnetization at 0 K, $T_g$=$g\mu_B H_{\mathrm{int}}/k_{\rm B}$ is the gap temperature, and  $z\left(\frac{3}{2},\frac{T_g}{T}\right)$ and $z\left(\frac{5}{2},\frac{T_g}{T}\right)$ are the correction terms which reduce to unity if the effective internal magnetic field vanishes. The two temperature terms above come from the harmonic ($k^2$) and anharmonic ($k^4$) terms in the spin-wave dispersion relation [Eq. (\ref{disp})]. For simplicity, disregarding anharmonicity and the gap corrections, Eq. (\ref{spin}) reduces to the so-called Bloch's $T^{3/2}$ law: 
\begin{equation} \label{Bloch}
M(T)=M(0)(1-BT^{3/2}).
\end{equation}
In conventional spin-wave theory the spin wave parameter $B$ and the spin wave stiffness coefficient $D$ are related through
\begin{equation} \label{stiffness}
D = \frac{k_{\rm B}}{4\pi} \left(\frac{2.612 \, g\mu_B}{M(0)B}\right)^{2/3} \; .
\end{equation}

The thermal demagnetization process of ferromagnetic metals at low temperatures ($T$$\ll$$T_{\rm C}$) can be explained by both localized\cite{14} and itinerant\cite{15} models. In the localized model, with static interactions between spins associated with localized electrons on atomic sites, spin waves correspond to coherent superposition of local spin deviations in the ferromagnetic state, and their equilibrium number density at finite temperature according to the Bose-Einstein distribution function yields the magnetization reduction according to Eq. (\ref{spin}). In the itinerant model, electrons move in the average field of other electrons/ions, and effective inter-site spin couplings are generated by exchange of the particle-hole propagator, strong correlation effects in which have been investigated recently using a systematic non-perturbative expansion scheme.\cite{spandey} The presence of thermally excited long wavelength spin waves in the ferromagnetic state at low temperature allows virtual excitation of majority-spin electrons to the minority-spin band due to electron-magnon coupling, resulting in spectral-weight transfer and consequent reduction of the magnetization $\langle n_{\uparrow} \rangle - \langle n_{\downarrow} \rangle$.    

 
\begin{table*} [t]
\centering
\caption{Magnetization vs. temperature data are fitted to Eq. (\ref{spin}) and the coefficients $B$, $C$ along with normalized $\chi^2$, R$^2$, $T_{\rm C}$, and $D$ are tabulated below.}
\begin{tabular}{l c c c c c c c c}
\hline
\hline
Sample  & M(0) & B & C & B$_{3/2}$ & $T_{\rm C}$ & normalized $\chi^2$ & R$^2$ & D \\
Composition & (emu) & (10$^{-4}$ K$^{-3/2}$) & (10$^{-6}$ K$^{-5/2}$) & & (K) & (10$^{-7}$) & & meV$\mathring{A}^2$ \\
\hline
Ni$_{81.0}$Fe$_{16.7}$Mo$_{2.3}$ & 3.33323$\pm$0.00005 & 0.1189$\pm$0.0004 & - & 0.23 & 720 & 0.099 & 0.99843 & 219 \\
 & 3.33286$\pm$0.00004 & 0.094$\pm$0.001 & 0.023$\pm$0.001 &  &  & 0.036 & 0.99941 & \\
Ni$_{80.4}$Fe$_{12.9}$Mo$_{6.7}$ & 1.4606$\pm$0.0002 & 0.294$\pm$0.005 & - & 0.32 & 495 & 2.91 & 0.97166 & 144 \\
 & 1.4601$\pm$0.0003 & 0.24$\pm$0.03 & 0.1$\pm$0.06 &  &  & 0.036 & 0.99941 & \\
Ni$_{83.4}$Fe$_{10.7}$Mo$_{5.9}$ & 0.79083$\pm$0.00006 & 0.345$\pm$0.003 & - & 0.35 & 470 & 2.08 & 0.99064 & 155 \\
 & 0.79038$\pm$0.00007 & 0.24$\pm$0.01 & 0.19$\pm$0.02 &  &  & 1.28 & 0.99439 & \\
Ni$_{83.5}$Fe$_{7.6}$Mo$_{8.9}$ & 3.13870$\pm$0.00002 & 0.864$\pm$0.007 & - & 0.49 & 320 & 3.31 & 0.99981 & 110 \\
Ni$_{83.1}$Fe$_{6.0}$Mo$_{10.9}$ & 1.20957$\pm$0.00005 & 3.68$\pm$0.01 & - & 0.90 & 182 & 0.75 & 0.99916 & 45 \\
Ni$_{83.2}$Fe$_{3.3}$Mo$_{13.5}$ & 0.2770$\pm$0.0001 & 42.7$\pm$0.1 & - & 1.29 & 45 & 63.9 & 0.99871 & 20 \\
 & 0.2755$\pm$0.0001 & 36.2$\pm$0.5 & 0.44$\pm$0.03 &  &  & 27.7 & 0.99944 & \\
\ \\
Ni$_{78.9}$Fe$_{18.1}$W$_{3.0}$ & 4.0269$\pm$0.0001 & 0.0982$\pm$0.0007 & - & 0.21 & 775 & 0.52 & 0.99332 & 255 \\
Ni$_{79.4}$Fe$_{14.1}$W$_{6.5}$ & 1.25119$\pm$0.00002 & 0.302$\pm$0.001 & - & 0.37 & 530 & 0.15 & 0.99945 & 133 \\
 & 1.25098$\pm$0.00002 & 0.275$\pm$0.002 & 0.040$\pm$0.002 &  &  & 0.07 & 0.99976 & \\
Ni$_{81.0}$Fe$_{11.8}$W$_{7.2}$ & 1.10312$\pm$0.00002 & 0.359$\pm$0.001 & - & 0.42 & 515 & 0.31 & 0.99921 & 137 \\
 & 1.10280$\pm$0.00001 & 0.311$\pm$0.002 & 0.070$\pm$0.002 &  &  & 0.04 & 0.9999 & \\
Ni$_{83.5}$Fe$_{7.6}$W$_{8.9}$ & 0.56687$\pm$0.00001 & 0.796$\pm$0.001 & - & 0.41 & 300 & 0.47 & 0.99956 & 120 \\
Ni$_{82.6}$Fe$_{6.9}$W$_{10.5}$ & 1.66087$\pm$0.00003 & 2.906$\pm$0.004 & - & 0.77 & 191 & 0.22 & 0.99963 & 53 \\
Ni$_{86.6}$Fe$_{3.1}$W$_{10.3}$ & 0.54141$\pm$0.00001 & 19.6$\pm$0.1 & - & 0.8 & 55 & 18.08 & 0.99825 & 28 \\
\hline
\label{taba}
\end{tabular}
\end{table*}

The low-temperature magnetization was measured from 2 K to 0.1 $T_{\rm C}$ or till 15 K at $\sim$ 6000 Oe for all the samples which is above their saturation fields of 1000-2000 Oe, although sample Mo13.5 did not actually saturate even at 6000 Oe due to its low-temperature mixed ferro-spin-glass phase. Figure 1 is a plot of $M$ vs. $T$ for $T \ll T_{\rm C}$. The solid lines are the best-fitted curves as indicated in the legends of the figures. The gap correction [Eq. ({\ref{spin})] was tried but it was found to be negligible since the gap temperature came out to be less than 1 K.

From Fig. 1 as well as Table I we find that the fits for some of the 12 samples that we have studied improve significantly if we include the $T^{5/2}$ term in Eq. (\ref{spin}). Not only the values of $\chi^2$ are smaller by a factor of 2-3 ($\sim$8 for W7.2), the deviation of the best-fitted curve from the data vs. $T$ is random as against systematic when we consider only the $T^{3/2}$ term. For the other samples the improvements are insignificant. We must note that for higher $T_{\rm C}$ samples (low Mo and W), $\Delta M/M$ is only 1$\%$ and only very high resolution SQUID measurements are able to isolate the anharmonic term in the magnon dispersion relation.

The $first$ significant feature emerging from the present investigation is the order of magnitude of the spin-wave parameter $B$. These values are found to be strongly enhanced with increasing Fe dilution in both the series of alloys, and are about 10 to 100 times larger than those found for conventional 3d ferromagnets (for bulk Fe it is $3.4 \times 10^{-6}$ K$^{-3/2}$). This sharp increase in the values of $B$ with increasing Fe dilution reflects an enhancement in the density of low-energy magnetic excitations due to weakening of the ferromagnetic couplings between the bulk Fe spins forming the percolating ferromagnetic matrix. The spin-wave stiffness constants are also correspondingly reduced, as seen in Table I.

\subsection*{C.	Intermediate-temperature behaviour}
In order to allow a qualitative comparison between the magnetization behaviour of different ferromagnetic systems over a much broader temperature scale extending nearly up to $T_{\rm C}$, it is convenient to use the normalized coefficient $B_{3/2}$ defined through the relation:\cite{16,17}
\begin{equation} \label{B3hlf}
\frac{M_s(0)-M_s(T)}{M_s(0)}= B_{3/2}\left(\frac{T}{T_C}\right)^{3/2} \ 
\end{equation}
in terms of the low-temperature magnetization (Eq. \ref{Bloch}), which yields $B_{3/2}$=$B T_{\rm C} ^{3/2}$. Significantly, this reduced coefficient $B_{3/2}$ provides an effective measure of the overall shape of the magnetization decay. For crystalline ferromagnets, where the magnetization falls sharply near $T_{\rm C}$, one obtains $B_{3/2} \approx 0.2$, irrespective of the Curie temperature. On the other hand, for reference, if the magnetization were to fall off as $M(0)(1-BT^{3/2})$ nearly upto $T_{\rm C}$, then clearly $B_{3/2} \approx 1$. Table \ref{taba} shows that with increasing Fe dilution, the $B_{3/2}$ values for both series of alloys systematically increase from $\sim 0.2$ to $\sim 1$, indicating that the magnetization decay near $T_{\rm C}$ becomes significantly slower in the Fe-poor alloys. This is the $second$ significant result which emerges from the intermediate-temperature regime.

Both these distinctive features of macroscopic magnetization behaviour observed in Ni-Fe-Mo and Ni-Fe-W alloys --- strong enhancement of $B$ with increasing Fe dilution and $B_{3/2}$ approaching unity --- have also been observed in squid magnetization studies of diluted magnetic semiconductors Ga$_{1-x}$Mn$_{x}$As,\cite{7} where they were ascribed to two distinct energy scales involved in the local spin dynamics associated with the formation of spin clusters. A brief review of the theoretical analysis of finite-temperature spin dynamics within a minimal model for diluted magnets will be helpful in understanding the observed magnetization behaviour.

Theoretical investigations in diluted magnets do yield competing ferromagnetic and antiferromagnetic interactions.\cite{17} Within a minimal model for diluted magnets involving spin-S localized impurity spins and host band fermions (carriers), magnetic interactions between two impurity spins at lattice sites $i$ and $j$ were calculated from $J_{ij}$=$J^2 (2S) \phi_{ij}$ in terms of the particle-hole propagator $\phi_{ij}$ evaluated in the ferromagnetic state. For the same impurity separation, the calculated magnetic couplings were found to exhibit a broad distribution, implying that the coupling between two impurity spins is not simply a function of their separation, but actually depends on the whole disorder configuration, suggesting shades of a complex system.

The broad distribution in the calculated ferromagnetic spin couplings was ascribed to the formation of small impurity spin clusters due to positional magnetic disorder. The cluster spin couplings were found to be strongly enhanced due to preferential accumulation of carriers in impurity spin clusters, whereas the consequent depletion of carriers from the bulk resulted in weakened bulk spin couplings between the bulk spins forming the ferromagnetic matrix.\cite{8,9,10}

The effects of such a broad distribution of ferromagnetic spin couplings on finite temperature spin dynamics was recently investigated in diluted magnets using a locally self-consistent magnon renormalization scheme,\cite{8,9,10} in analogy with renormalized spin-wave theory in ordered ferromagnets.\cite{18} The local magnetization $\langle{S_i^z}\rangle$ of each individual impurity spin at site $i$ in a quantum spin-S disordered ferromagnet was obtained by self-consistently solving the three coupled equations:
\begin{equation} \label{savg}
\langle{S_i^z}\rangle=\frac{(S-\Phi_i)(1+\Phi_i)^{2S+1}+(S+1+\Phi_i)\Phi_i^{2S+1}}{(1+\Phi_i)^{2S+1}-\Phi_i^{2S+1}}
\; ,
\end{equation}
where the local site-dependent boson occupation numbers:
\begin{equation}
\Phi_i=\sum_{l}\frac{\vert{\phi_l^i}\vert^2}{e^{\beta\omega_l}-1}
\end{equation}
in terms of the renormalized magnon energy eigenvalues $\omega_l$ and eigenfunctions $\phi_l$ obtained from the renormalized magnon Hamiltonian:
\begin{equation}
{\cal{H}}_{ij}=\sqrt{2\langle{S_i^z}\rangle}\left(J^2[\chi^0]_{ij}\right)\sqrt{2\langle{S_j^z}\rangle} \; .
\end{equation}
The locally self-consistent magnetization $\langle{S_i^z}\rangle$ thus obtained clearly showed rapid thermal demagnetization and nearly paramagnetic behaviour of the weakly coupled bulk (FM matrix) spins, whereas the strongly coupled cluster spins resist thermal demagnetization and thus prolong the magnetic order near $T_{\rm C}$. The overall picture from the averaged magnetization was that while the ferromagnetic $T_{\rm C}$ was suppressed by positional-disorder-induced spin cluster formation, the magnetization decay was distinctly stretched near $T_{\rm C}$.

These distinctive features of microscopic spin dynamics behaviour in a generic diluted magnet were shown to quantitatively affect the two readily accessible characteristics $B$ and $B_{3/2}$ of macroscopic magnetization behaviour. While the calculated spin-wave parameter $B$ was found to be sharply enhanced with increasing dilution [Fig. 2 (upper panel)] due to weakened bulk spin couplings and softening of low-energy spin excitations, the stretching of magnetic order near $T_{\rm C}$ due to strongly coupled cluster spins resulted in enhanced $B_{3/2}$$\sim$1, indicating slower thermal demagnetization approximately as $T^{3/2}$ (dashed line) over a much broader temperature range, as shown for five different disorder configurations (lower panel). In sharp contrast, the ordered ferromagnet yielded, for the same dilution and carrier concentration, a conventional magnetization decay with $B_{3/2} \approx 0.2$ and $T^{3/2}$ behaviour (dashed line) only in the low-temperature regime. Here the given notation refers to impurity concentration ($x$), carrier concentration ($p$), bandwidth ($W$), and host-impurity energy offset ($\epsilon_d$).  

\begin{figure} 
\centering
\includegraphics[width=8cm]{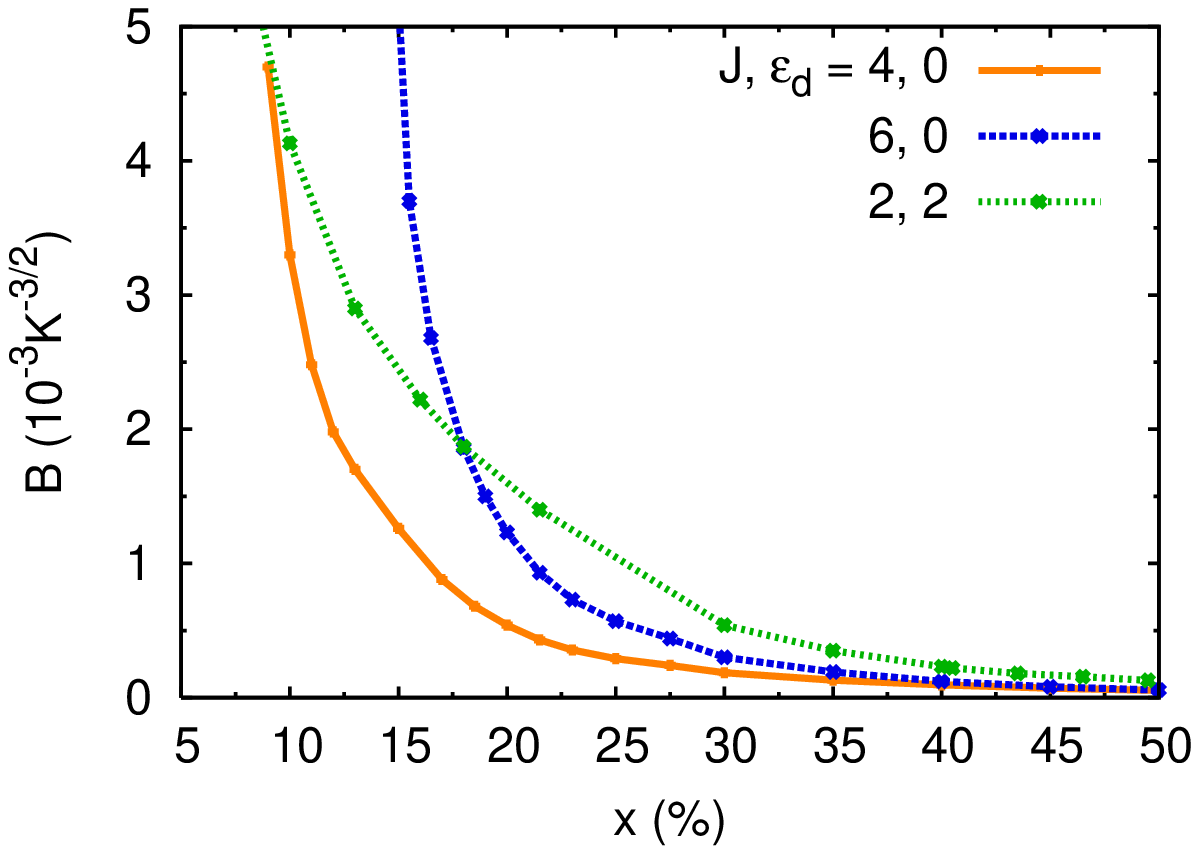}
\ \\
\includegraphics[width=8cm]{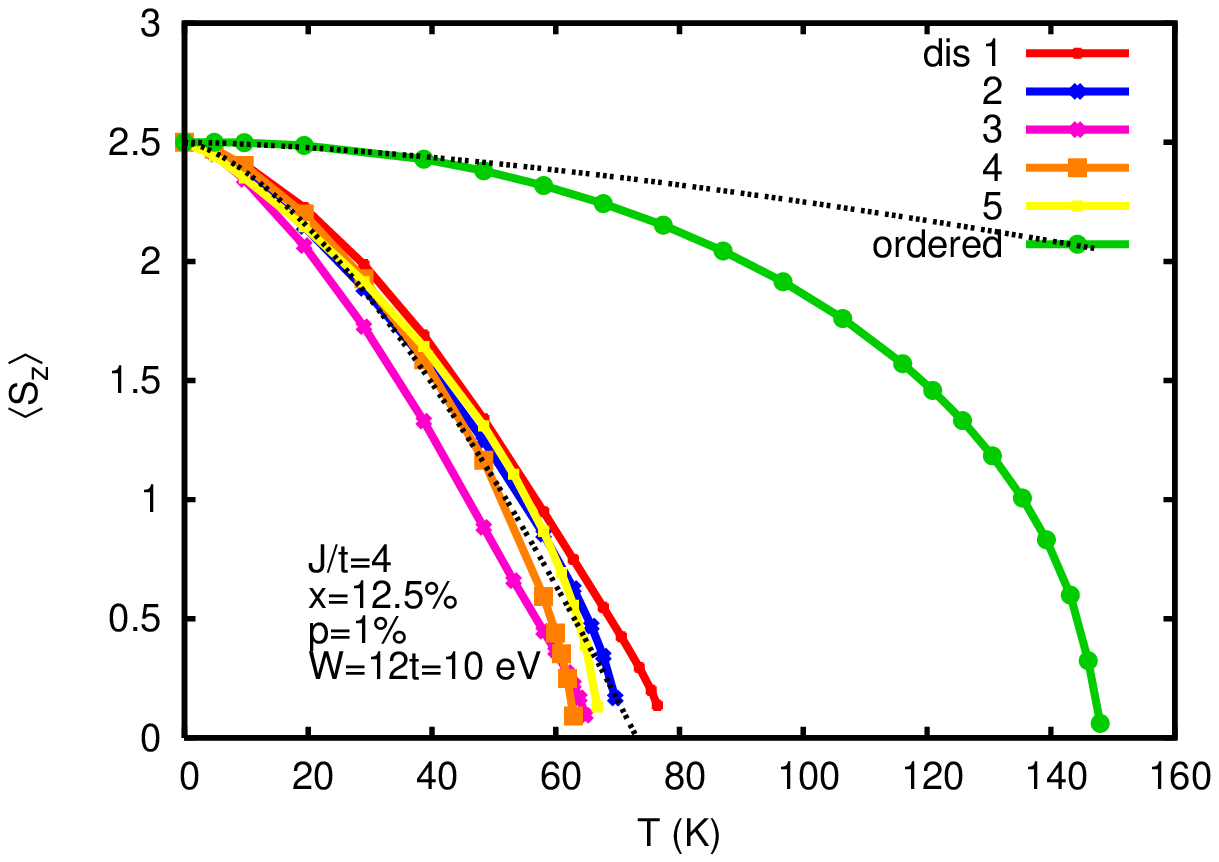}
\caption{(Color Online) Rapid enhancement of the calculated spin-wave parameter $B$ with dilution (upper panel), and (lower panel) nearly $T^{3/2}$ fall-off of site-averaged magnetization and stretching of magnetic order near $T_{\rm C}$ (from ref. [7]).}
\label{figtheory}
\end{figure}

As shown in Table \ref{taba}, a very similar behaviour is obtained for the two magnetization (spin-wave) coefficients $B$ and $B_{3/2}$ from our finite-temperature magnetization investigation of Ni-Fe-Mo and Ni-Fe-W alloys. While the $B$ values are sharply enhanced with increasing Fe dilution (the spin stiffness is correspondingly sharply suppressed), the $B_{3/2}$ values approach 1 with increasing Fe dilution in both alloys, as compared to about 0.2 for ordered ferromagnets. Similar enhancements were reported recently from SQUID magnetization measurements on the diluted magnetic semiconductors $\rm Ga_{1-x}Mn_xAs$,\cite{7} and in earlier studies on amorphous ferromagnetic alloys in comparison with crystalline ferromagnets.\cite{13,14}

\subsection*{D.	Magnetic phase transition and the associated critical exponents}
In the present work, we have also studied the critical exponents associated with the magnetic phase transition of the alloys containing more than 8 at.\% of Mo/W. The specific reason to study the critical exponents and the amplitudes is to see the consistency of our findings with respect to the above magnetization data analysis. In the last section we have seen evidence of spin clusters with relatively strongly coupled spins in the alloys with dearth of Fe content. How does this spin clustering affect the critical exponents and critical amplitudes? As a first step, ac susceptibility has been measured in all alloys with a temperature increment of 0.2 K or less so that their $T_{\rm C}$ could be found accurately within $\pm$ 0.1 K. The magnetization was measured in fields ranging from 0 to 20 kOe at various temperatures around 2\% of $T_{\rm C}$ on either side. 

The critical exponent $\beta$ associated with $M_s$, the spontaneous magnetization, is given by 
$M_s$=$B\vert\varepsilon\vert^\beta$ for $T$$<$$T_C$, where $\varepsilon$=$(T-T_{\rm C})/T_{\rm C}$ and $B$ is the corresponding critical amplitude. The critical exponent $\gamma$ is related to $\chi_0$, the zero-field dc susceptibility, through the equation $\chi_0^{-1}$=$\Gamma^{-1}\vert\varepsilon\vert^\gamma$, where $\Gamma^{-1}$ is the critical amplitude. The magnetic field dependence of $M$ at $T_{\rm C}$ gives us the third critical exponent $\delta$ through the relation $M$=$DH^{1/\delta}$, $D$ being the critical amplitude. All these three exponents follow a static scaling relationship, $\delta$=$1 + \gamma/\beta$. In order to find these exponents one has to know the transition temperatures very accurately.
First the isothermal magnetization $M$ was plotted as in a simple Arrott plot ($M^2$ vs. $H/M$) which uses the mean-field exponents. However, when they did not give linear isotherms, the modified Arrott-Noakes (AN) plot was tried, i.e., $M^{1/\beta}$ versus $(H/M)^{1/\gamma}$. 

To plot the AN isotherms one has to know the values of $\beta$ and $\gamma$ beforehand. Instead of guessing the initial values we have calculated $\beta$ and $\gamma$ using a simple FORTRAN code where we applied the method of least squares fitting. We have calculated the average chi-square ($\chi^2$) for all possible combinations of $\beta$ and $\gamma$, where $\beta$ was varied between 0.2 and 0.6 in steps of 0.001 and $\gamma$ from 0.9 to 1.5 in similar steps. Then the combination of $\beta$ and $\gamma$, for which the average $\chi^2$ is a minimum and the slopes of all the straight lines are almost equal, i.e., the set of isotherms are truly parallel to each other, are taken. Amazingly, the extrapolated isotherm at $T_{\rm C}$ indeed passes through the origin making us quite confident of the plots. Then using those values of $\beta$ and $\gamma$ we have plotted in Fig. \ref{fig2} the modified Arrott-Noakes (AN) plot, a set of isotherms for 10 \% Mo alloy, the y-intercepts of these straight lines give $M_s$ and the x-intercepts $\chi_0^{-1}$.  Figure \ref{fig2} (upper panel), a typical AN plot, also gives the values of $\beta$ and $\gamma$ for the best-fitted isotherms. Although $\gamma$ is close to the mean-field value of 1, $\beta$ is much lower than the mean-field value of 0.5. Deviations of the values of $\beta$ and $\gamma$ from the mean-field values have been found both in many magnetic glasses as well as in crystalline ferromagnets.  

\begin{figure} 
\centering
\includegraphics[width=12cm,height=10cm]{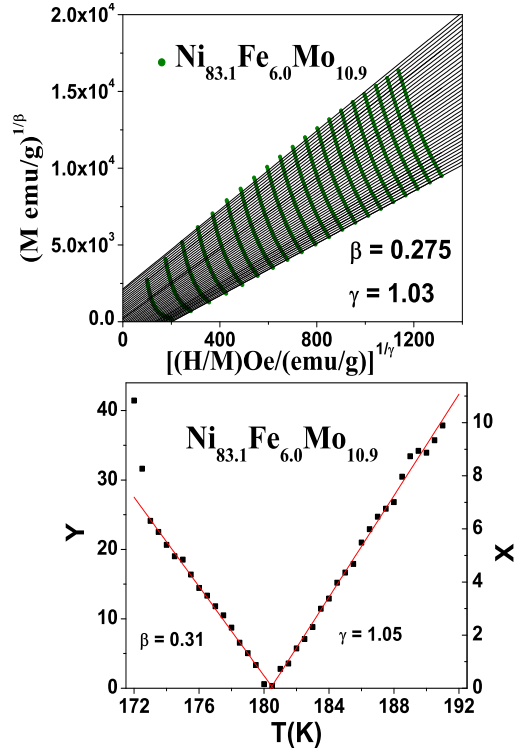}
\caption{(Color Online) The modified Arrott-Noakes (AN) plot and X (T) and Y (T) vs. temperature plots for Mo10.9 alloy with values of $\beta$ and $\gamma$ for which the best-fitted isotherms are obtained.}
\label{fig2}
\end{figure}

There is another way of calculating $\beta$ and $\gamma$, that is using the Kouvel-Fisher method. The set of equations used in this method are:
\begin{equation} 
Y(T)=\left(\frac{d\mathrm{ln} M_s}{d\mathrm{T}}\right)^{-1}=\left(\frac{1}{\beta}\right)(T-T_C) 
\end{equation}
and
\begin{equation} 
X(T)=\left(\frac{d\mathrm{ln} \chi_0^{-1}}{d\mathrm{T}}\right)=\left(\frac{1}{\gamma}\right)(T-T_C) .
\end{equation}
These equations are valid only in the critical region where temperatures are very near $T_{\rm C}$. Here we have used the $M_s$ and $\chi_0^{-1}$ values from the intercepts of the AN plots. In the critical region $Y(T)$ vs. $T$ and $X(T)$ vs. $T$ are both straight lines with slopes of (1/$\beta$) and (1/$\gamma$), respectively. In this method a priori knowledge of $T_{\rm C}$ is not needed and if the results are consistent, then $Y(T)$ and $X(T)$ will intersect the $T$ axis at the same point. Figure \ref{fig2} (lower panel) also shows the typical $X(T)$ and $Y(T)$ plots for 10 \% Mo alloy. Both of them intersect the $T$-axis at $T$=$T_{\rm C}$=180.5 K.

We have used both the methods for all our samples to check the consistency of the results. Except for Mo13.5 alloy, all the other samples gave quite reasonable values of the critical exponents. However, it is to be noted that unlike other alloys, Mo13.5 has a reentrant spin-glass phase (RSG) phase below $T_g$=10 K ($T_{\rm C}$ is $\sim$ 44 K) which might possibly affect the critical behavior since the transition may not be a pure ferro-para one although we used $M(T,H)$ data only from 
$T$=43-45 K, far away from 10 K.  Nevertheless, our calculated values of the critical exponents from the two methods did not match at all. More importantly, the value of $\beta$ was above and that of  $\gamma$ was below the mean-field values. This is quite unphysical as the mean-field case is the limiting one, since it is a rather crude theory. Presence of more than one phase due to improper homogenization and/or final annealing might cause such deviations in the critical exponents. Table II below gives the values of $T_{\rm C}$, $\beta$, $\gamma$, and  $\delta$. The critical exponent values suggest that Mo8.9 and W8.9 almost follow the mean-field model whereas the others, except Mo13.5 are close to 3D-Ising rather than that of pure nickel, which follows the 3D-Heisenberg model.

\begin{table} [t]
\centering
\caption{Alloy compositions, values of $T_{\rm C}$ and critical exponents, obtained both experimentally and from KF analysis along with those for pure Ni and those from existing theories.}
\begin{tabular}{l c c c c c c c c }
\hline
\hline
Sample & \multicolumn{2}{ c }{$T_{\rm C}$} & \multicolumn{2}{ c }{$\beta$} & \multicolumn{2}{ c }{$\gamma$} & \multicolumn{2}{ c }{$\delta$} \\
Composition & Expt. & KF & Expt. & KF & Expt. & KF & Expt. & 1+$\frac{\beta}{\gamma}$ \\
\hline
Ni$_{83.5}$Fe$_{7.6}$Mo$_{8.9}$ & 316.0 & 316.4 & 0.48 & 0.42 & 1.21 & 1.31 & 3.54 & 3.50 \\
Ni$_{83.1}$Fe$_{6.0}$Mo$_{10.9}$ & 180.5 & 180.4 & 0.275 & 0.31 & 1.03 & 1.05 & 4.68 & 4.74 \\
Ni$_{83.2}$Fe$_{3.3}$Mo$_{13.5}$ & 44.55 & 46.2 & 0.59 & 0.54 & 0.92 & 0.88 & 2.22 & 1.64 \\
Ni$_{83.5}$Fe$_{7.6}$W$_{8.9}$ & 305.0 & 300.0 & 0.43 & 0.45 & 1.30 & 1.37 & 4.22 & 4.05 \\
Ni$_{82.6}$Fe$_{6.9}$W$_{10.5}$ & 191.0 & 190.0 & 0.34 & 0.39 & 1.23 & 1.19 & 4.67 & 4.64 \\
Ni$_{86.6}$Fe$_{3.1}$W$_{10.3}$ & 58.15 & 58.15 & 0.33 & 0.35 & 1.304 & 1.29 & 4.93 & 4.54 \\
\hline
Ni$^a$ & 627.4 & & 0.378 & & 1.34 & & 4.58 & 4.54 \\
\hline
Mean-field &  & & 0.50 & & 1.00 & & 3.00 &  \\
3D-Ising &  & & 0.312 & & 1.25 & & 5.00 &  \\
3D-Heisenberg &  & & 0.378 & & 1.405 & & 4.76 &  \\
\hline
$^a$ Reference 21.
\label{tabb}
\end{tabular}
\end{table}

\begin{table} [b]
\centering
\caption{The critical amplitudes of the alloys and those of Ni for comparison.}
\begin{tabular}{l c c c}
\hline
\hline
Sample & $\Gamma^{-1}$ & B$_0$ & D$_0$ \\
composition & (kOe-g/emu) & (emu/g) & \\
\hline
Ni$_{83.5}$Fe$_{7.6}$Mo$_{8.9}$ & 8.4 & 21.4 & 7.36 \\
Ni$_{83.1}$Fe$_{6.0}$Mo$_{10.9}$ & 4.2 & 20.0 & 5.14 \\
Ni$_{83.2}$Fe$_{3.3}$Mo$_{13.5}$ & 1.7 & 2.5 & 1.9 \\
Ni$_{83.5}$Fe$_{7.6}$W$_{8.9}$ & 8.7 & 27.2 & 7.27 \\
Ni$_{82.6}$Fe$_{6.9}$W$_{10.5}$ & 6.1 & 25.9 & 5.13 \\
Ni$_{86.6}$Fe$_{3.1}$W$_{10.3}$ & 4.6 & 8.8 & 4.79 \\
\hline
Ni$^a$ & 19 & 83 & 30 \\
\hline
$^a$ Reference 21.
\label{tabc}
\end{tabular}
\end{table}

We have also calculated the critical amplitudes since they are important for a complete knowledge of the critical behavior near ferro-para transition. The magnetization above and below $T_{\rm C}$ satisfies a single scaling equation given by 
$m$=$f_\pm (h)$ where $m$=$\vert\varepsilon\vert^{-\beta} M(\varepsilon,H)$ and $h$=$\vert\varepsilon\vert^{-\beta\delta}H$ are called scaled magnetization and the scaled magnetic field. The above relations show that $m$ as a function of $h$ falls on two different universal curves $f_-(h)$ for $T$$<$$T_{\rm C}$ and $f_+(h)$ for $T$$>$$T_{\rm C}$. If the values of the critical exponents found here are correct, then all the data will fall on two distinct curves confirming their correct choice. The critical amplitudes $B_0$=$m_0$, $\Gamma^{-1}$=$h_0/m_0$ and $D_0$ are obtained from the intercepts of the $\ln$-$\ln$ plot of $M_s$ vs. $\vert\varepsilon\vert$, $\chi_0^{-1}$ vs. $\vert\varepsilon\vert$ and $M$ vs. $H(\varepsilon=0)$, respectively. Figure \ref{fig3} shows, respectively  $\ln \chi_0^{-1}$ vs. $\ln \varepsilon$ plot for W10.3, ln $M_s$ vs. $\ln H$ for W10.5, $\ln M_s$ vs. $\ln \varepsilon$ for Mo10.9, and $M/ \vert\varepsilon\vert^\beta$ vs. $H/\vert\varepsilon\vert^{\beta\delta}$ for W10.3 including the calculated values of the corresponding critical amplitudes. Figure \ref{fig3} (lower right panel) shows clearly the correctness of our procedure. Table \ref{tabc} gives the critical amplitudes of some of the samples. We observe that with the increasing Fe dilution, the critical amplitudes $\Gamma^{-1}$, $B_0$, and $D_0$ decrease systematically and all three have values lower than those for Ni. This is the $third$ observable consequence of strongly coupled cluster spins, indicating significantly reduced participation by the bulk spins forming the ferromagnetic matrix in the critical behaviour. 

\begin{figure} [t]
\centering
\includegraphics[width=8cm,height=7cm]{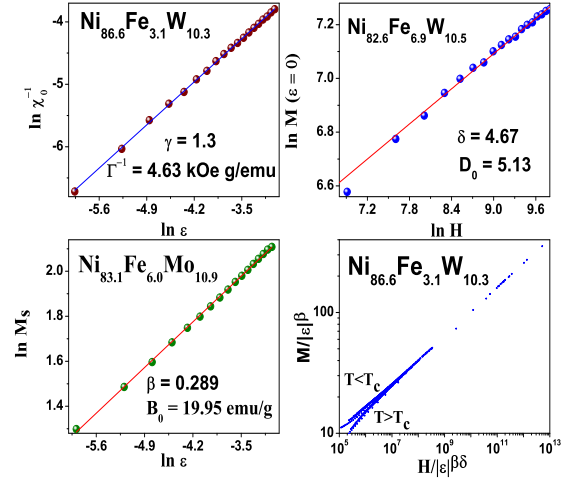}
\caption{(Color Online) . $\chi_0^{-1}$ vs. $\ln \varepsilon$ plot for  W10.3 alloy, $\ln M_s$ vs. $\ln H$ plot for W10.5 alloy, $\ln M_s$ vs. $\ln \varepsilon$ plot for Mo10.9 alloy and $M/\vert\varepsilon\vert^\beta$ vs. $H/\vert\varepsilon\vert^{\beta\delta}$ plot for W10.3 alloy.}
\label{fig3}
\end{figure}

The locally self-consistent spin dynamics analysis shows that as the temperature falls below $T_{\rm C}$, the cluster spins rapidly get magnetized.\cite{8,9,10} A similar behaviour is expected above $T_{\rm C}$ in an external magnetic field. Thus, the cluster spins contribute dominantly to the high magnetic susceptibility of the alloys near $T_{\rm C}$, and hence result in a suppression of $\chi_0 ^{-1}$ and of the critical amplitudes. 

Particularly for the Mo13.5 sample, the extremely low measured values of critical amplitudes along with the unusual critical exponent values (see Table II) indicates dominant finite-cluster contribution just below the percolation threshold, with no diverging spin correlation length and no true critical behaviour. This is consistent with the emergence of spin-glass behaviour at this composition due to frustrated cluster spins locked in frozen non-collinear orientations which also results in smaller spontaneous magnetization along the z-direction. 

\section{Conclusion}
In conclusion, we find that all three macroscopic magnetization characteristics  --- the spin-wave parameter $B$, the reduced coefficient $B_{3/2}$, and the critical amplitudes $\Gamma^{-1}$ --- corresponding to the low, intermediate, and the critical temperature regimes respectively, yield distinctive spin dynamics signatures associated with strongly coupled cluster spins in these alloys. 

The dilution behaviour of the two magnetization coefficients extracted from the macroscopic magnetization behaviour of finite-temperature spin dynamics in the two alloy systems is indicative of a broad distribution of magnetic spin interactions between the magnetically active Fe atoms. The sharp enhancement in the measured spin-wave parameter $B$ accompanying the spin stiffness reduction with Fe dilution clearly indicates weakened bulk spin couplings and softening of low-energy spin-wave modes. Furthermore, the magnitude of the reduced coefficient $B_{3/2}$ rapidly approaches 1 with increasing Fe dilution, and the slower magnetization decay with temperature indicates presence of strongly coupled cluster spins which resist thermal demagnetization and stretch the magnetic order near $T_{\rm C}$. 

As also observed in diluted magnetic semiconductors and metallic glasses, this sharp enhancement with dilution is thus suggestive of the two magnetization coefficients as universal macroscopic indicators of spin clustering and disorder-induced distribution of magnetic interactions in disordered ferromagnets. The reduction of the measured critical amplitudes with Fe dilution due to significantly reduced participation by the bulk spins forming the ferromagnetic matrix in the critical behaviour is consistent with this picture. Similar suppression of critical amplitudes was observed in $\rm Fe_x Ni_{80-x} P_{14} B_6 $ alloys,\cite{20} and was ascribed to the growth of spin clusters as $T$ approaches $T_{\rm C}$ and the consequent reduced participation of remaining bulk (FM matrix) spins in the FM-PM phase transition. 

Except for the Mo13.5 alloy, which has a spin-glass phase below $T_g$=10 K, all the other samples gave quite reasonable values of the critical exponents. Non-universal values of critical exponents with $\beta \approx 0.55$ have also been found in recent experimental studies\cite{haetinger_2009} of the ferromagnetic transition in spin-glass re-entrant metallic alloys such as $\rm Au_{0.81}Fe_{0.19}$.

\section*{Acknowledgements}
MB acknowledges financial support from UGC, Govt. of India.

\end{document}